\documentclass{optica-article}

\journal{opticajournal} % for journals or Optica Open

\articletype{Research Article}

\usepackage{lineno}
\usepackage{graphicx,physics,siunitx,comment,booktabs}

\begin{document}

\title{Range Emulator: A Compact Paraxial Optical System to Emulate Long-Distance Monochromatic Laser Propagation}

\author{SUBARU SHIBAI\authormark{1,*} and KIWAMU IZUMI,\authormark{2}}

\address{\authormark{1}Department of Physics, University of Tokyo, Bunkyo, Tokyo 113-0033, Japan\\
\authormark{2}Department of Space Astronomy and Astrophysics, Institute of Space and Astronautical Science, JAXA\\
}

\email{\authormark{*}s-s-shibai0301@g.ecc.u-tokyo.ac.jp} %% email address is required; see note below about the corresponding author designation

% use {asbstract*} to suppress the copyright line. Copyright information will be added in production.

\begin{abstract*} 
Emulating long-distance light propagation on a laboratory scale is essential for the ground-based testing of intersatellite optical systems. To address this challenge, we propose and analyze an optical system called the Range Emulator (RE) to reproduce the spatial propagation effects of a long-distance beam within a compact apparatus. Our analysis identifies the three-lens solutions that implement the RE. Through a numerical exploration, we quantify the fundamental trade-off between system compactness and manufacturing precision. This work provides a practical framework for designing compact optical testbeds for future multi-satellite laser link technologies.
\end{abstract*}

%%%%%%%%%%%%%%%%%%%%%%%%%%  body  %%%%%%%%%%%%%%%%%%%%%%%%%%
\section{Introduction}
Light propagation provides a powerful platform for analog signal processing that enables filtering, storing, and amplifying the signal. In this context, the optical path length constitutes a fundamental resource. Classic examples include multiple reflectors in the Michelson-Morley experiment~\cite{michelsonRelativeMotionEarth1887} and the Herriott delay line~\cite{herriottFoldedOpticalDelay1965}. In recent years, a prime example may be ground-based gravitational wave detectors, which utilize kilometer-long optical paths to detect spacetime distortions \cite{aasiAdvancedLigo2015,acerneseAdvancedVirgoSecondgeneration2014,akutsuOverviewKAGRADetector2021}. The feasible arm length of such terrestrial facilities is typically limited by geographical constraints and construction costs. More fundamentally, their sensitivity below \SI{1}{Hz} is precluded by seismic and Newtonian noise.

Space lifts both restrictions and extends the available optical path to millions of kilometers. Such baselines also relax the displacement sensitivity required to reach a given strain. Therefore, space-based gravitational wave detectors deploying intersatellite laser interferometer have been proposed
\cite{amaro-seoaneLaserInterferometerSpace2017,kawamuraJapaneseSpaceGravitational2011b,ruanTaijiProgramGravitationalwave2020,luoTianQinSpaceborneGravitational2016}. These missions are crucial for observing sub-\SI{1}{Hz} gravitational waves. Observing astronomical sources such as massive black hole binaries, extreme mass-ratio inspirals, and the stochastic gravitational wave background from the early universe, these detectors are expected to open a new window to the universe. The success of these missions depends on the establishment of precise laser links over distances from thousands to millions of kilometers.

To address this unprecedented challenge, the SILVIA mission was recently proposed \cite{itoSILVIAUltraprecisionFormation2025}. This mission aims to demonstrate ultra-precise formation flying and laser interferometry using three satellites separated by \SI{100}{m}. For intersatellite interferometry, it is essential to establish alignment algorithms to compensate for spatial propagation effects, specifically beam position shifts and size changes.
Developing such technologies requires a ground-based testbed. However, building a testbed for intersatellite optical links poses significant difficulties. Physically separated platforms on the ground suffer from uncorrelated environmental disturbances, such as seismic motion and temperature fluctuations. Furthermore, atmospheric turbulence degrades optical performance, yet constructing long vacuum tubes to mitigate it may not be affordable. Consequently, a compact optical system capable of reproducing the spatial propagation effects of long distances has been sought.

In this paper, we propose an optical system, the Range Emulator (RE), designed to emulate the spatial propagation effects of light traveling for a long distance within a few meters. Our primary achievement is the discovery of a three-lens solution. Furthermore, we demonstrate that this three-lens solution possesses inherent design freedom. This freedom can be strategically exploited to enhance the practical feasibility of the RE by optimizing robustness against manufacturing tolerances. To efficiently navigate the multi-dimensional design space, we introduce a numerical methodology based on probabilistic solution sampling. This approach generates a diverse catalog of high-performance configurations, providing a comprehensive overview of the design landscape and its inherent trade-offs, a particularly valuable feature for a system without established design heritage such as the RE.

Unlike previous optical simulators developed for space-based laser communications~\cite{inagakiFreespaceSimulatorLaser1991,wanOngroundSimulationOptical2010b}, the RE adopts a fundamentally different optical design approach. While those earlier systems relied on Fourier optics to reproduce far-field diffraction patterns, our RE is specifically engineered to emulate the Rayleigh range propagation of a Gaussian beam. This unique capability is essential for validating high-precision interferometry missions like SILVIA.

A related class of devices, known as spaceplates, has recently been developed to reproduce the transfer function of free-space propagation within a reduced physical length~\cite{guoSqueezeFreeSpace2020,reshefOpticReplaceSpace2021,shastriWhatExtentCan2022,sorensenLargescaleOpticalCompression2023}. These devices are nonlocal flat optics that reproduce the propagation phase over the wide angular range required for imaging applications. The RE, in contrast, is intended for intersatellite metrology and needs to accept only a narrow angular range of about \SI{10}{\milli\radian}, which allows a far higher compression of the propagation distance. It also operates at a different scale, emulating propagation over \SI{100}{m} within a few meters using a single Gaussian mode, and is accordingly described by Ray Transfer Matrix.

The paper is organized as follows. Section~\ref{sec:range_emulator} introduces the concept of RE and formulates it as a Ray Transfer Matrix (RTM) synthesis problem. Section~\ref{sec:method} details our numerical design strategy. Section~\ref{sec:results} presents the discovery of the three-lens solutions and quantitatively evaluates their feasibility. Section~\ref{sec:discussion} discusses practical limitations and implementation considerations. Finally, Section~\ref{sec:conclusion} summarizes our findings and outlines future directions.

\section{Range Emulator}\label{sec:range_emulator}
First we define the RE and formulate it as a constrained RTM synthesis problem. Subsequently, we review the previous studies on the RTM decomposition which are potentially applicable to the RE design.

\subsection{The definition}
\begin{figure}[t]
    \centering
    \includegraphics[width=\textwidth]{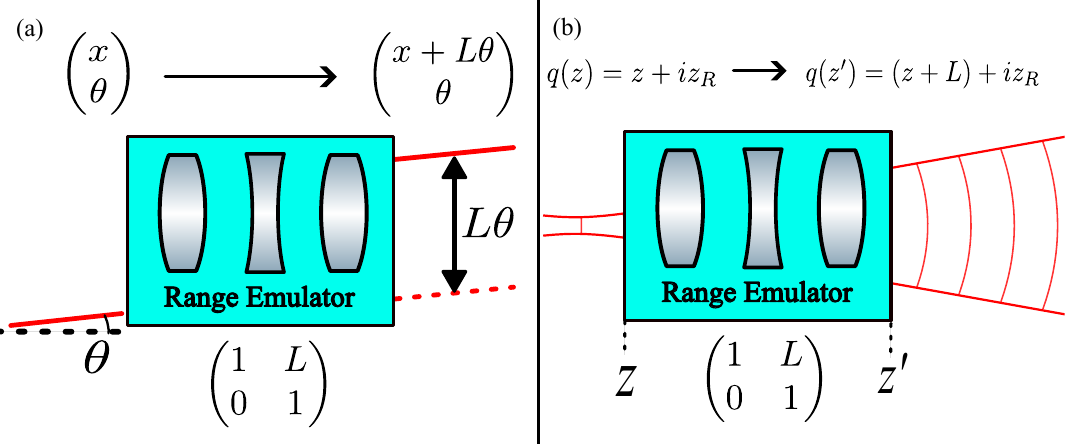}
    \caption{Conceptual image of the Range Emulator (RE). (a) The RE mimics geometric propagation effects, transforming the position and angle of an incoming beam as if it had traveled a long distance. (b) The RE also replicates the evolution of Gaussian beam parameters, such as beam radius and wavefront curvature, corresponding to long-distance propagation.
    }
    \label{fig:RS_image}
\end{figure}

 The RE must be able to replicate two key physical effects. First, it must transform the beam position and angle as if it has propagated a given long distance. Second, it must transform the Gaussian beam parameters accordingly. Fortunately, these two properties can be fully described by the Ray Transfer Matrix (RTM) as illustrated in Fig.~\ref{fig:RS_image}. It should be noted that the RE does not emulate the time-of-flight delay, which can instead be readily implemented by electronic means.

The physical effects given above can be represented in the matrix form. According to the standard formulation~\cite{siegman1986lasers}, we introduce $\mathrm{L}(D)$ as the RTM for a thin lens with diopter $D$. Similarly, we let $\mathrm{P}(d)$ be the RTM of free space propagations over a distance $d$. They are expressed as
\begin{equation}
\mathrm{P}(d)=\begin{pmatrix}
        1&d\\
        0&1
    \end{pmatrix}
    \quad \textrm{and }\quad 
    \mathrm{L}(D)=
    \begin{pmatrix}
        1&0\\
        -D&1
    \end{pmatrix}.
\end{equation}

We hereafter assume the optical systems to be constructed by a combination of lenses and propagation in free space as shown in Eq.\eqref{RS_equation}
\begin{equation}
    \begin{pmatrix}
        1&L\\
        0&1\\
    \end{pmatrix}
    =\mathrm{L}(D_1)\mathrm{P}(d_1)\mathrm{L}(D_2)\dots \mathrm{P}(d_{n-1})\mathrm{L}(D_n).
\label{RS_equation}
\end{equation}

To ensure the practical utility of the RE, we place the following constraint on the total length,

\begin{equation}
        \sum_i d_i \leq L_{\mathrm{max}} \ll L.
\label{RS_constraint}
\end{equation}

Here, $L$ is a propagation distance to be emulated, and $L_{\mathrm{max}}$ is the maximum feasible size of the apparatus, typically on the order of meters for a single optical bench. 

\subsection{Previous studies on RTM synthesis}
We now wish to confirm whether there are parameter combinations $(d_1,D_1,\dots,d_{n-1}, D_n)$ that satisfy Eq.~\eqref{RS_equation} under the constraint Eq.~\eqref{RS_constraint}. This question pertains to a fundamental problem of RTM synthesis. It is known that three lenses are sufficient to realize an arbitrary $2\times 2$ RTM system \cite{01081985,liuMinimalOpticalDecomposition2008}. However, these proofs assume that lens optical powers and propagation distances could take any unconstrained value. Indeed, if the total length is not constrained, the problem becomes trivial because the target matrix $\mathrm{P}(L)$ itself is the solution. This defeats the purpose of the RE. It is therefore necessary to find solutions that adhere to practical upper and lower bounds on all system parameters. We cannot rely on the existing general proofs and must instead seek specific solutions that satisfy these constraints.

Previous studies on the RTM have provided several useful matrix decomposition forms that can be applied to the RE. Simon and Wolf~\cite{simonStructureSetParaxial2000b} provided a concrete solution for the RE,
\begin{equation}
    \mathrm{P}(L)=\mathrm{R}\qty(\frac{\pi}{2})\mathrm{L}(L)\mathrm{R}\qty(-\frac{\pi}{2}),\label{Simon_solution}
\end{equation}
where $\mathrm{R}(\theta)$ is referred to as a phase space rotator defined as
\begin{equation}
    \mathrm{R}(\theta)=
    \begin{pmatrix}
        \cos{\theta}&\sin{\theta}\\
        -\sin{\theta}&\cos{\theta}\\
    \end{pmatrix}.
\end{equation}
 The implementation of the phase space rotator requires a single lens for the phase in $0<\theta< \pi$, whereas it requires two lenses for $-\pi <\theta<0$. Consequently, four lenses are necessary in total for the practical construction for Eq.~(\ref{Simon_solution}).

If an inversion of the transverse coordinate is acceptable, a three-lens solution exists,
\begin{equation}
    -\mathrm{P}\left(L\right)=\mathrm{R}\qty(\frac{\pi}{2})\mathrm{L}(L)\mathrm{R}\qty(\frac{\pi}{2}).
\end{equation}
An equivalent three-lens configuration has also been obtained in~\cite{sorensenLargescaleOpticalCompression2023} using Fourier optics.

Similarly, the other decomposition~\cite{nazarathyFirstorderOpticsCanonical1982} gives another solution as
\begin{equation}
    \mathrm{P}\left(L\right)=\mathrm{L}\qty(-\frac{1}{L})\mathrm{M}\qty(-L)\mathrm{R}\qty(-\frac{\pi}{2})\mathrm{L}\qty(-\frac{1}{L}),
\end{equation}
where $\mathrm{M}(\alpha)$ is a magnifier matrix, 
\begin{equation}
    \mathrm{M}(\alpha)=\begin{pmatrix}
        \alpha & 0\\
        0&\alpha^{-1}
    \end{pmatrix},
\end{equation}
with $\alpha$ a magnification factor.
The magnifier matrix requires at least two lenses, resulting in six lenses in total.

In summary, the preceding studies reveal that while analytical solutions for the RE exist, they are not guaranteed to be minimal or practical. To investigate the most practical solution for RE and evaluate its feasibility, we depart from analytical decomposition and instead adopt a numerical approach.

\section{Numerical Exploration}\label{sec:method}
\subsection{Overview}
We perform a numerical exploration to identify practical solutions. The objective is not only to find solutions that theoretically realize the RE functions, but also to quantitatively evaluate their implementation feasibility.

To avoid conflating physical correctness and engineering preferences, we classify the design criteria into two hierarchical cost functions, namely, the primary cost function and the secondary cost functions.
The primary cost function $\mathrm{C}_{\mathrm{p}}$ represents the fundamental requirements derived from the RTM formulation in Eq.~\eqref{RS_equation}, quantifying how closely a given optical system approximates the desired behavior. The ensemble of solutions must satisfy a predefined threshold ($\mathrm{C}_{\mathrm{p}} < \epsilon$) to realize the RE. In contrast, the secondary cost function represents practical implementation metrics. In our case, the total system length $\mathrm{C}_{\mathrm{L}}$ and robustness against manufacturing tolerances $\mathrm{C}_{\mathrm{R}}$ will be incorporated. These are used to evaluate the quality of the valid candidates post-hoc.

\begin{figure}[t]
    \centering
    \includegraphics[width=\textwidth]{Solution_valley.pdf}
    \caption{Schematic illustrations of the solution exploration. (a) Genetic Algorithm (GA) explores the solution space to find initial candidates that satisfy the target performance ($\mathrm{C}_{\mathrm{p}} < \epsilon$). (b) Hamiltonian Monte Carlo (HMC) samples the parameter space around these candidates, generating a diverse ensemble of solutions. (c) The secondary cost functions ($\mathrm{C}_{\mathrm{L}}$ and $\mathrm{C}_{\mathrm{R}}$) are calculated for each sample, allowing for the evaluation of practical feasibility.}
    \label{fig:solution_valley}
\end{figure}

Based on this classification, the exploration strategy is structured into three main phases as illustrated in Fig.~\ref{fig:solution_valley}. The phases are:
\begin{enumerate}
    \item \textbf{Global Search --} 
    First, we employ a Genetic Algorithm (GA) to perform a global search minimizing $\mathrm{C}_{\mathrm{p}}$. The goal of this phase is to find one of the possible solutions that satisfy the primary optical requirements. This step is crucial because gradient-based or local sampling methods often fail to converge without a good initial guess in the non-convex landscape of optical design.
    
    \item \textbf{Local Solution Sampling --} 
    Starting from the solution found in the previous phase, we generate a diverse ensemble of solutions using a probabilistic sampling method based on Hamiltonian Monte Carlo (HMC). Because the thick-lens model introduces corrections that are not captured by the closed-form thin-lens solution manifold, the valid region must be explored numerically. This step explores the "solution valley" of designs that satisfy the primary threshold ($\mathrm{C}_{\mathrm{p}} < \epsilon$), providing a comprehensive map of valid optical configurations.
    
    \item \textbf{Feasibility Analysis --} 
    Finally, for each solution in the generated ensemble, we evaluate the secondary cost functions ($\mathrm{C}_{\mathrm{L}}$ and $\mathrm{C}_{\mathrm{R}}$). By analyzing the trade-offs between system size and robustness, we identify the most practical design candidates.
\end{enumerate}

Since phases 1 and 2 are implementation details of the search itself, they are described in Appendix~\ref{sec:appendix}. In the following, we define the model and the primary cost function, and then describe the feasibility analysis of Phase 3.

\subsection{Model and Definition of Primary Cost Function}
For the numerical exploration, we adopt specifications relevant to SILVIA. We set the target propagation distance to be $L=\SI{100}{m}$ and the maximum apparatus size $L_{\mathrm{max}}=\SI{3}{m}$. The target RTM is therefore given by
\begin{equation}
    \begin{pmatrix}
        A_0 & B_0 \\
        C_0 & D_0
    \end{pmatrix}=
    \begin{pmatrix}
        1 &100  \\
        0 & 1
    \end{pmatrix}.
\end{equation}

To model the RE optical system, we employ a thick lens model, assuming symmetric bi-convex or bi-concave lenses for simplicity. The choice of thick lenses is motivated by the fact that analytical solutions for the RE often require high-diopter lenses, whose performance cannot be accurately described by the simple thin lens model. The optical system is characterized by the parameter vector $\mathbf{p}$, comprising lens curvatures $R_i$, thicknesses $t_i$, and spacings $d_i$, as 
\begin{equation}
    \mathbf{p}=\qty( R_1, t_1, d_1, R_2, \dots, d_{n-1}, R_n, t_n)^T.
\label{parameter_vector}
\end{equation}
All lenses are assumed to be made of fused silica, with a refractive index of $n = 1.444$ at the operating wavelength of $\lambda = \SI{1550}{nm}$.

This parameter vector $\mathbf{p}$ is subject to the following three practical constraints. First, the minimum absolute radius of curvature ($\abs{R_i}$) shall be $\SI{5}{mm}$. Second, the maximum lens thickness $t_i$ shall be $\SI{20}{mm}$. Finally, any individual propagation distance $d_i$ must not exceed $\SI{3}{m}$.

The primary cost function, $\mathrm{C_{\mathrm{p}}}(\mathbf{p})$, evaluates the deviation of the designed system from the target RTM and ideal mode-matching efficiency $\eta$. It is defined as follows:
\begin{equation}
    \mathrm{C_\mathrm{p}}(\mathbf{p})= \qty(\frac{\mathrm{A}(\mathbf{p})-A_{0}}{\Delta A})^2+\qty(\frac{\mathrm{B}(\mathbf{p})-B_{0}}{\Delta B})^2+\qty(\frac{\mathrm{C}(\mathbf{p})-C_{0}}{\Delta C})^2+\qty(\frac{\mathrm{D}(\mathbf{p})-D_{0}}{\Delta D})^2+\qty(\frac{\eta(\mathbf{p})-\eta_0}{\Delta \eta})^2.\label{cost_function}
\end{equation}
Here, the terms $\Delta A,\Delta B,\Delta C,\Delta D$, and $\Delta \eta$ are the acceptable tolerances for each respective parameter, serving as normalization factors. In our investigation, we set $\Delta A = \Delta D = 0.01$, $\Delta B = \SI{0.1}{m}$, $\Delta C = \SI{0.01}{m^{-1}}$, and $\Delta \eta = 0.01$.
Although this definition of the primary cost function might seem redundant since three parameters are sufficient for the RTM, this formulation better reflects the design intent by automatically prioritizing the most stringent condition.

The term $\eta(\mathbf{p})$ represents the Mode-Matching Efficiency (MME) with $\eta_0=1$ set for the target value. 
The MME is defined as
\begin{equation}
    \eta = \frac{\int \abs{E_1(\mathbf{r})^*E_2(\mathbf{r})\dd{\mathbf{r}}}^2}{\int \abs{E_1(\mathbf{r})}^2\dd{\mathbf{r}}\int \abs{E_2(\mathbf{r})}^2\dd{\mathbf{r}}},
\end{equation}
where $E_1(\mathbf{r})$ is the complex electric field distribution of the beam propagated through the ideal free space $L$, and $E_2(\mathbf{r})$ is that of the beam propagated through the RE system. The asterisk denotes the complex conjugate.

The MME is evaluated at a wavelength of \SI{1550}{nm}. The input beam is specified by the complex beam parameter $q_{\mathrm{in}} = (-50 + 50 i) \si{m}$, corresponding to a waist radius of $w_0 \cong \SI{5.0}{mm}$ located at the midpoint of the emulated \SI{100}{m} path.

\subsection{Feasibility Analysis}
After generating the solution ensemble (Appendix~\ref{sec:appendix}), we perform a post-hoc analysis on the solution ensemble to evaluate its practical metrics. In this analysis, we focus on two key metrics as follows. The first metric is the total physical length, which is calculated as the sum of all propagation distances. Minimizing this metric reflects the primary desire for developing a compact system and reveals the physical limits of the RE design.
\begin{equation}
    \mathrm{C}_{\mathrm{L}}(\mathbf{p}) = \sum_i d_i.
\end{equation}

The second metric is the robustness against manufacturing and alignment errors. This metric quantifies the system's sensitivity to small perturbations in its parameters. While there are various ways to define robustness \cite{moritzgohlerRobustnessMetricsConsolidating2016}, we define robustness cost function as the maximum normalized sensitivity of the system's performance metrics to fractional parameter perturbations, such that a smaller value indicates lower sensitivity to manufacturing errors. The metric is calculated as follows:
\begin{equation}
    \mathrm{C}_{\mathrm{R}}(\mathbf{p})=\mathrm{max}\qty[\frac{\sqrt{\sum_{i} \qty(\pdv{X}{p_i} p_i)^2}}{\Delta X}: X\in\qty{A,B,C,D,\eta}].
\end{equation}\label{eq:robustness}
The reciprocal of this cost function, $1/\mathrm{C}_\mathrm{R}(\mathbf{p})$, corresponds to the required fractional accuracy of the system parameters.

After computing these two metrics for the entire ensemble, designers can visualize the results (e.g., by plotting robustness versus total length) to understand the inherent trade-offs. To interpret the high-dimensional data, Principal Component Analysis (PCA) can be applied. This allows for the identification of the key design parameters that are the primary drivers of system robustness, enabling an informed and practical final design selection.

\section{Results}\label{sec:results}
\subsection{Numerical Exploration Results}
Following the procedure outlined in the previous section, we performed the numerical explorations by varying the number of lenses. We confirmed that a three-lens configuration is the minimum required to realize the RE numerically. The local sampling yielded 206,391 valid solutions that satisfy the primary threshold (Appendix~\ref{sec:appendix}).
\begin{figure}[t]
    \centering
    \includegraphics[width=\textwidth]{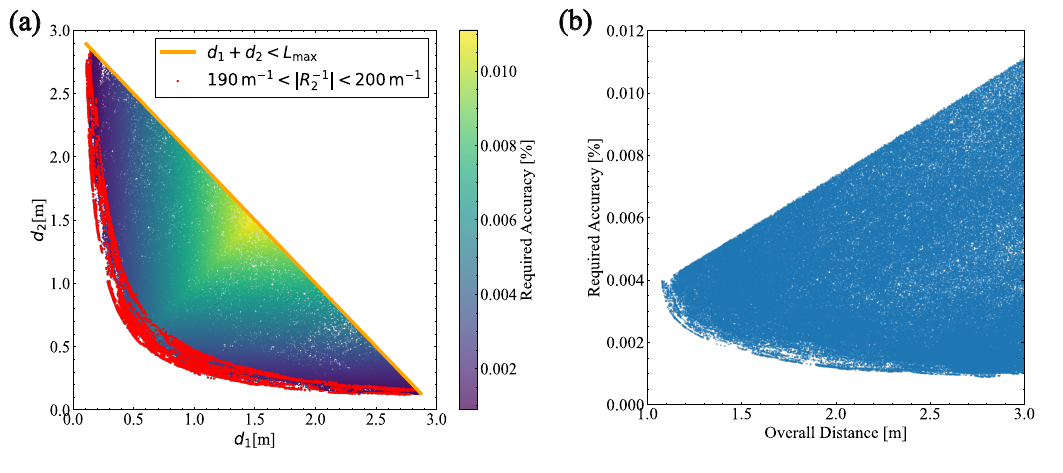}
    \caption{(a) Visualization of the HMC sampling results in the parameter space of lens distances $d_1$ and $d_2$. Color indicates the fractional accuracy required to meet the performance of the RE, given by $1/\mathrm{C}_{\mathrm{R}}$; a larger value corresponds to a looser requirement.
    The parameter's range is limited by the maximum size of apparatus $L_{\mathrm{max}}=\SI{3}{m}$ and the curvature limitation of the center lens ($R_2$). (b) The required manufacturing accuracy is plotted against the total system length ($d_1+d_2$). There are inherent trade-offs between compactness and tolerance to fabrication errors.}
    \label{fig:hmc_sampling}
\end{figure}

Fig.~\ref{fig:hmc_sampling}(a) shows the ensemble of collected samples plotted in the parameter space of the lens distances $d_1$ and $d_2$. All points shown satisfy the primary system requirements. The boundary is determined by the maximum apparatus size $C_\textrm{L}=d_1+d_2$ and the curvature limit of the center lens, which is constrained by the minimum radius of curvature $\abs{R_2}>\SI{5}{mm}$.

Fig.~\ref{fig:hmc_sampling}(b) presents the results of the sensitivity analysis, plotting the required manufacturing accuracy $1/\mathrm{C}_{\mathrm{R}}$ for each sample against the total system length. The plot reveals a trade-off: systems with a shorter total length tend to be more sensitive to inaccuracies in the parameters, while more robust solutions require a longer configuration.

\begin{figure}[tb]
    \centering
    \includegraphics[width=0.5\textwidth]{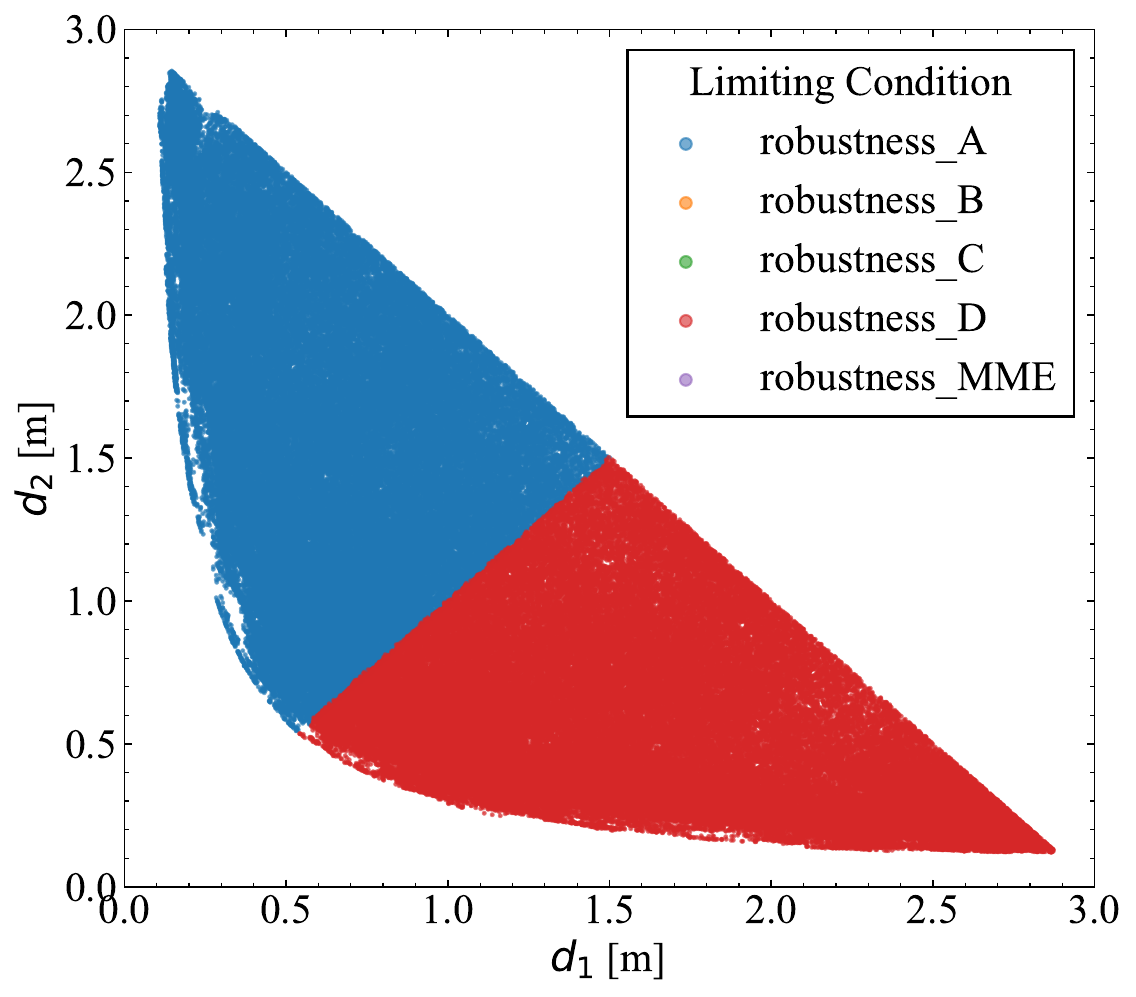}
    \caption{Color indicates which of the performance metrics $\qty{A,B,C,D,\eta}$ imposes the tightest requirement, i.e., which term attains the maximum in Eq.~\eqref{eq:robustness}, at each point in the ($d_1,d_2$) plane.}
    \label{fig:limiting_condition}
\end{figure}

Fig.~\ref{fig:limiting_condition} shows which component (A,B,C,D, MME) decides the required accuracy. The limiting condition of the RE is determined by A or D, and it depends on the balancing of the $d_1$ and $d_2$. The most robust solutions lie on the boundary between the regions where $A$ and $D$ are limiting, which corresponds to the symmetric condition $d_1 \cong d_2$.

From this large ensemble, a designer can select any solution according to the requirements at hand. As a recommended configuration for the RE, we propose a solution that takes the total optical length $d_1 + d_2$ close to its allowed maximum and satisfies $d_1 \cong d_2$, so that the required accuracy is as loose as possible. The corresponding parameter vector and the achieved RTM elements are listed in Tables~\ref{tab:design} and \ref{tab:performance}.

\begin{table}[t]
\centering
\caption{Parameters of the recommended three-lens configuration and the
fractional accuracy required for each parameter. All lenses are symmetric bi-convex or bi-concave fused-silica elements ($n=1.444$). Here the requirement is evaluated for each parameter individually, in contrast to the aggregate metric of Eq.~\eqref{eq:robustness}, in order to identify which parameter is binding. The last column
indicates which of the performance metrics $X\in \{ A,B,C,D,\eta \}$ imposes the tightest
requirement on that parameter.}
\label{tab:design}
\begin{tabular}{lrrc}
\toprule
Parameter & Value [\si{mm}] & Required accuracy [\%] & Limiting $X$ \\
\midrule
$R_1$ & 1369.26 & 0.015 & $A$ \\
$t_1$ &   16.47 & 8.0   & $A$ \\
$d_1$ & 1487.19 & 0.016 & $A$ \\
$R_2$ &  -23.07 & 0.093 & $B$ \\
$t_2$ &   18.02 & 0.90  & $B$ \\
$d_2$ & 1495.07 & 0.016 & $D$ \\
$R_3$ & 1374.12 & 0.015 & $D$ \\
$t_3$ &    2.06 & 65    & $D$ \\
\bottomrule
\end{tabular}
\end{table}

\begin{table}[t]
\centering
\caption{Residual Error of Ray transfer matrix elements and mode-matching efficiency
achieved by the recommended configuration.}
\label{tab:performance}
\begin{tabular}{ccccc}
\toprule
$A-A_0$ & $B-B_0$ [\si{m}] & $C-C_0$ [\si{m^{-1}}] & $D-D_0$ & $\eta-\eta_0$ \\
\midrule
 \num{-9.4e-5}&  \num{1.2e-3}&  \num{-2.458e-7}& \num{6.9e-5}&  \num{-1.7e-9}\\
\bottomrule
\end{tabular}
\end{table}

\subsection{Three-Lens Solution of Range Emulator}
We have numerically demonstrated the existence of a three-lens solution for the 100-\si{m} RE and, for completeness, note that no solution exists for fewer lenses, consistent with known RTM-synthesis results~\cite{01081985,liuMinimalOpticalDecomposition2008}.

The case with $n=1$ is trivial. For the one with $n=2$, we can expand the right-hand side of Eq. \eqref{RS_equation} as
\begin{equation}
\mathrm{L}(D_1)\mathrm{P}(d_1)\mathrm{L}(D_2) =
\begin{pmatrix}
1-d_1D_2 & d_1\\
-D_1-D_2(1-d_1D_1) & 1-d_1D_1
\end{pmatrix}.
\end{equation}
For this matrix to equal the target propagation matrix $\mathrm{P}(L)$, the upper right element requires $d_1=L$. This in fact reduces the system to the trivial solution where $D_1=D_2=0$, defeating the purpose of the RE. Therefore, a meaningful solution with two or fewer lenses is not achievable.

\begin{figure}[t]
    \centering
    \includegraphics[width=\textwidth]{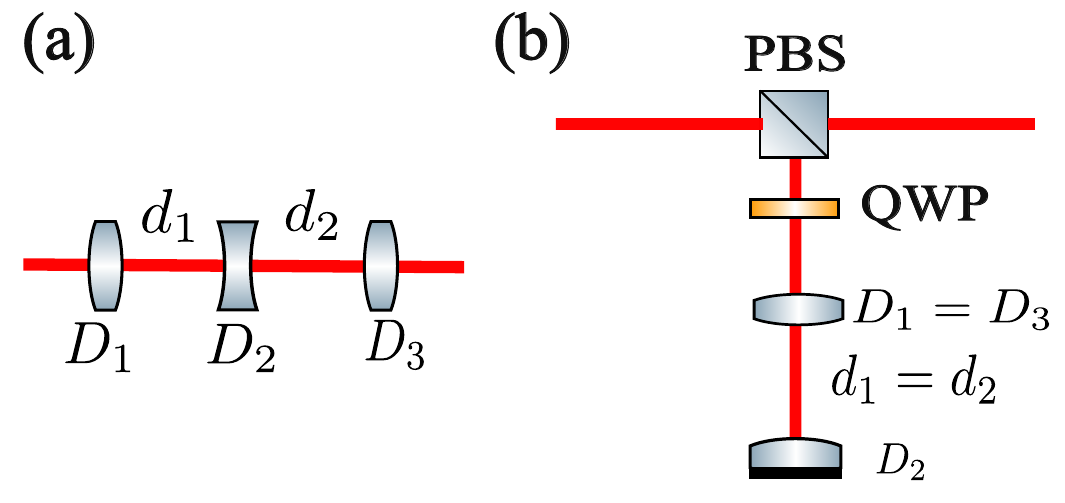}
    \caption{(a) schematic of the three-lens Range Emulator (RE) configuration. (b) A folded configuration of the RE that reduces the number of alignment parameters by enforcing symmetry, with $d_1=d_2$ and $D_1=D_3$. This design is particularly effective when only one polarization direction is used.}
    \label{fig:RS_configuration}
\end{figure}

The numerical exploration above employs the thick-lens model, which is necessary for accurate performance evaluation. However, to understand the structure of the solution distribution, it is useful to derive a closed-form expression.

Solving Eq.~\eqref{RS_equation} for $n=3$, we derived the following expressions for the lens diopters $D_1$, $D_2$, and $D_3$ in terms of the lens separation distances $d_1$ and $d_2$:
\begin{equation}
    \begin{split}
D_1 &= \frac{1}{d_1} - \frac{1}{L}\qty(\frac{d_2}{d_1}+1), \\
D_2 &= \frac{1}{d_1} + \frac{1}{d_2} - \frac{L}{d_1d_2}, \\
D_3 &= \frac{1}{d_2} - \frac{1}{L}\qty(\frac{d_1}{d_2}+1).
    \end{split}\label{RE_solution}
\end{equation}

Eq.~\eqref{RE_solution} indicates that the design of the RE has two degrees of freedom. While the selection of independent variables can be tailored to specific design goals, we have selected $d_1$ and $d_2$ as the primary variables. The choice is motivated by our requirement to construct a compact RE system; the remaining parameters ($D_1,D_2,D_3$) are subsequently determined as dependent variables. Since $d_1$,$d_2$ and $L$ must be positive, and under the condition where $L\gg d_1+d_2$, the diopters $D_1$ and $D_3$ are always positive i.e., convex lenses, while $D_2$ is negative i.e., concave lens. Fig.~\ref{fig:RS_configuration} (a) illustrates the convex-concave-convex configuration for the three-lens RE.

\section{Discussion}\label{sec:discussion}
\subsection{Limits of the Three-Lens RE}
Eq.~\eqref{RE_solution} indicates that a more compact system requires a central lens with a larger negative optical power. This implies that the maximum emulatable distance is ultimately constrained by the feasible system size $L_{\mathrm{max}}$ and the maximum achievable optical power of the concave lens $|D_{\mathrm{max}}|$. For a symmetric configuration where $d_1=d_2=L_{\mathrm{max}}/2$, the maximum emulatable distance $L_{\mathrm{limit}}$ can be estimated as
\begin{equation}
L_{\mathrm{limit}} =L_{\mathrm{max}}\qty(1+\frac{L_{\mathrm{max}}D_{\mathrm{max}}}{4}).
\end{equation}
Substituting our constraints of $L_{\mathrm{max}}=\SI{3}{m}$ and $D_{\mathrm{max}}=\SI{200}{m^{-1}}$ gives $L_{\mathrm{limit}}=\SI{453}{m}$, this value represents the theoretical upper bounds for a three-lens RE under these specific constraints.

To emulate distances beyond this limit, two approaches can be considered. A first approach of simply increasing the physical size of the apparatus or $L_{\mathrm{max}}$ is often impractical due to space and cost constraints. The other approach, which is more promising, is to increase the number of lenses. Adding lenses allows the high optical power required for long-distance emulation to be distributed among several elements. This strategy will be essential for emulating the kilometer-scale distances needed for future space-based optical missions \cite{amaro-seoaneLaserInterferometerSpace2017,kawamuraJapaneseSpaceGravitational2011b,ruanTaijiProgramGravitationalwave2020,luoTianQinSpaceborneGravitational2016}.

\subsection{Trade-offs and Practical Implementation}
As quantitatively shown in Fig.~\ref{fig:hmc_sampling}(b), the trade-off between compactness and robustness is a fundamental characteristic of the RE design. For our specific case with $L=100$~m, the most robust configuration within $\SI{3}{m}$ requires a parameter accuracy of approximately $0.01\%$. This level of accuracy can be achieved with current technology. However, the accuracy requirement will become increasingly stringent when emulating longer distances with more lenses, potentially reaching a point where implementation becomes impractical.

The demanding accuracy requirement motivates the adoption of a simplified practical implementation. A symmetric configuration ($d_1=d_2$, $D_1=D_3$), which can be physically realized by folding the optical path with a mirror as shown in Fig.~\ref{fig:RS_configuration}(b), provides two key benefits. First, it reduces the number of independent components that require precise alignment. Second, it simplifies performance verification; the symmetry guarantees that RTM diagonal elements $A$ and $D$ are equal, meaning the entire matrix can be characterized by measuring just two elements. It should be noted, however, that such folded designs are only suitable for a single polarization state. 

We note that the present analysis covers only the parameters contained in $\mathbf{p}$, namely the curvatures, thicknesses, and separations. Decentering and tilt of the individual lenses are not included, and they may dominate the error budget in a real assembly. Kinematic mounts offer adjustability but have limited absolute precision. Extending the parameter vector to include these degrees of freedom is left for future work; we are currently preparing an experimental setup for this purpose. Optical aberrations, which lie outside the RTM formalism, will be addressed in the same framework.

\subsection{Application of the RE}
While our study on the RE is meant specifically for the development of  intersatellite laser interferometry, its core concept holds potential for broader applications. Any measurement technique that benefits from a long optical path within a compact setup could leverage the RE design. For instance, the sensitivity of the optical lever technique, which converts a small angular displacement into a larger lateral shift after propagation, could be enhanced by using an RE.

Furthermore, the RE can emulate negative propagation distances. Setting $L < 0$ in Eq.~\eqref{RE_solution} reverses the sign of $D_2$, so that all three elements become convex lenses and the resulting RTM has $B = -|L|$. This inverts the evolution of the Gaussian beam parameter $q$. This capability broadens the design horizon for various precision laser optical systems.

\section{Conclusion}\label{sec:conclusion}
In this work, we introduced an optical system, the Range Emulator (RE), capable of emulating the spatial effects of long-distance light propagation within a compact apparatus. Our key contributions include the identification of a three-lens configuration and the derivation of a general analytical formula for this solution. Furthermore, our numerical exploration, based on a probabilistic sampling method, successfully mapped the design space for the RE with a target length of 100~m. This analysis quantitatively revealed the fundamental trade-off between system compactness and its robustness against manufacturing and alignment errors, providing a practical guide for implementation.

For future work, we plan to construct and verify an RE with the target length of 100~m based on the designs presented here, including an experimental evaluation of the misalignment sensitivities discussed above. Furthermore, we plan to extend our numerical framework to explore configurations with more than three lenses, aiming to find practical solutions for simulating the kilometer-scale distances required for next-generation intersatellite laser interferometer missions. Through these efforts, we aim to provide a powerful and flexible toolkit for the on-ground testing and validation of advanced optical systems.
\begin{backmatter}
\bmsection{Funding}
This research was supported by Forefront Physics and Mathematics Program to Drive Transformation (FoPM), a World-leading Innovative Graduate Study (WINGS) Program, the University of Tokyo.
\bmsection{Acknowledgement}
We thank Yutaro Enomoto for careful reading of the manuscript and useful suggestions. We also thank the SILVIA interferometer team members for the fruitful discussion.
\bmsection{Disclosures*}
The authors declare no conflicts of interest.
\bmsection{Data Availability*}
The data underlying the results presented in this paper are not publicly available at this time but may be obtained from the authors upon reasonable request.
\end{backmatter}

\appendix
\section{Implementation Details of the Numerical Exploration}\label{sec:appendix}
In this appendix, we describe the implementation details of the global search (Phase 1) and the local solution sampling (Phase 2) introduced in Sec.~\ref{sec:method}

\subsection{Global Search}

To obtain a suitable starting point for the HMC sampling, we first perform a global search. The optimization landscape in lens design is typically non-convex, often characterized by broad "solution valleys" where gradient-based methods are prone to stagnation in local minima~\cite{hoschelGeneticAlgorithmsLens2018,yangAutomaticLensDesign2022,sturlesiInvitedPaperFuture1991}. To mitigate this, we employ a Genetic Algorithm (GA), a global heuristic method robust against such complex landscapes~\cite{hoschelGeneticAlgorithmsLens2018,albuquerqueMultiobjectiveApproachOptimization2011,thibaultEvolutionaryAlgorithmsApplied2005}.

We implemented the GA using the \texttt{Platypus} library for Python~\cite{davidPlatypusFrameworkEvolutionary2024} to minimize the primary cost function (Eq.~\eqref{cost_function}) below the threshold $\epsilon_{\mathrm{p}}=0.01$. The algorithm was configured with a population size of 100 and run for 50,000 generations. We utilized tournament selection with size of 2, Simulated Binary Crossover (SBX) with the probability of 0.9, and polynomial mutation with the probability of 0.1. This global search successfully yielded a viable three-lens configuration, i.e., convex-concave-convex, which serves as the seed for the subsequent local sampling.

\subsection{Local Solution Sampling}

Following the global search, the next step is to explore the solution space locally to generate a diverse ensemble of valid configurations. We adopt a Markov Chain Monte Carlo (MCMC) method for this purpose, different from optimization techniques. While standard methods often aggregate multiple criteria into a single weighted cost function or aim for a Pareto front using multi-objective optimization algorithms, our goal is to sample the broad region satisfying the primary cost function threshold $\mathrm{C}_{\mathrm{p}} < \epsilon_{\mathrm{p}}$. Although this exhaustive sampling is computationally intensive, it yields a comprehensive understanding of the design landscape, enabling a post-hoc analysis of trade-offs between various secondary criteria without rerunning the optimization. This post-hoc approach has precedents in earlier design studies \cite{ivanssonMarkovchainMonteCarlo2014,mcguireDesigningEasilyManufactured2006}.

In this study, we implemented Hamiltonian Monte Carlo (HMC) ~\cite{nealMCMCUsingHamiltonian2011} using the \texttt{NumPyro} library~\cite{phan2019composable}. We define the target probability distribution for the parameter vector $\mathbf{p}$ based on the primary cost function, assuming a uniform prior over the permissible range:
\begin{equation}
P(\mathbf{p}) \propto \exp(-\beta \mathrm{C}_{\mathrm{p}}(\mathbf{p})).\label{likelihood}
\end{equation}
Here, $\mathrm{C}_{\mathrm{p}}(\mathbf{p})$ serves as the potential energy for the Hamiltonian dynamics. The hyperparameter $\beta$ is analogous to an inverse temperature that controls the concentration of the samples.  While a smaller $\beta$ allows for broader exploration, a value that is too low can cause the sampler to accept high-error solutions. We selected $\beta=0.1$ as providing a good balance.

The HMC sampling process was initialized using the solution discovered by the GA. We utilized the No-U-Turn Sampler (NUTS) \cite{hoffmanNoUTurnSamplerAdaptively2014}, an efficient variant of HMC. The process involved 2,000 warm-up iterations, followed by the collection of $5\times 10^6$ samples. Finally, we filtered this ensemble, retaining only the solutions that satisfied the primary cost function threshold $\mathrm{C}_{\mathrm{p}}(\mathbf{p})<\epsilon_{\mathrm{p}}$. A total of 206,391 samples met this criterion.

%%%%%%%%%%%%%%%%%%%%%%% References %%%%%%%%%%%%%%%%%%%%%%%%%

%%%%%%%%%% If using BibTeX:
\bibliography{reference}

\end{document}